\documentclass[5p,twocolumn]{elsarticle}
\usepackage{graphicx,latexsym}
\usepackage{dcolumn}
\usepackage{amssymb,amsmath,bm}
\usepackage{subfigure}
\usepackage{braket}
\usepackage{siunitx}
\usepackage{array}
\usepackage{float}
\usepackage[nopar]{lipsum}
\usepackage{caption}
\usepackage{multirow}
\usepackage{bigstrut}

\biboptions{sort&compress}

\usepackage[super]{nth}

\newcommand{\angstrom}{\text{\normalfont\AA}}
\usepackage{hyperref}
\hypersetup{
    pdfnewwindow=true,       
    colorlinks=true,         
    linkcolor=blue,          
    citecolor=blue,          
    filecolor=magenta,       
    urlcolor=black           
}

\usepackage[normalem]{ulem}

\def\sec#1{Sec.\ \ref{#1}}

\def\fig#1{Fig.\ \ref{#1}}

\journal{}

\begin{document}

\begin{frontmatter}



\title{Exploring electronic, optical, and phononic properties of MgX (X=C, N, and O) monolayers using first principle calculations}

\address[a1]{Division of Computational Nanoscience, Physics Department, College of Science,
	\\ University of Sulaimani, Sulaimani 46001, Kurdistan Region, Iraq}
\address[a2]{Computer Engineering Department, College of Engineering,
	\\ Komar University of Science and Technology, Sulaimani 46001, Kurdistan Region, Iraq}
\author[a1,a2]{Nzar Rauf Abdullah}
\ead{nzar.r.abdullah@gmail.com or nzar.abdullah@univsul.edu.iq}

\author[a3]{Botan Jawdat Abdullah}
\address[a3]{Physics Department, College of Science- Salahaddin University-Erbil, Erbil 44001, Kurdistan Region, Iraq}

\author[a4]{Yousif Hussein Azeez}
\address[a4]{Physics Department, College of Science, University of Halabja, Kurdistan Region, Iraq}

\author[a5]{Vidar Gudmundsson}
\address[a5]{Science Institute, University of Iceland, Dunhaga 3, IS-107 Reykjavik, Iceland}

\begin{abstract}

The electronic, the thermal, and the optical properties of hexagonal MgX monolayers (where X=C, N, and O) are investigated via first principles studies. Ab-initio molecular dynamic, AIMD, simulations using NVT ensembles are performed to check the thermodynamic stability of the monolayers. We find that an MgO monolayer has semiconductor properties with a good thermodynamic stability, while the MgC and the MgN monolayers have metallic characters. The calculated phonon band structures of all the three considered monolayers shows no imaginary nonphysical
frequencies, thus indicating that they all have excellent dynamic stability.
The MgO monolayer has a larger heat capacity then the MgC and the MgN monolayers. The metallic monolayers demonstrate optical response in the IR as a consequence of the metal properties, whereas the semiconducting MgO monolayer demonstrates an active optical response in the near-UV region.
The optical response in the near-UV is beneficial for nanoelectronics and photoelectric applications. A semiconducting monolayer is a great choice for thermal management applications since its thermal properties are more attractive than those of the metallic monolayer in terms of heat capacity, which is related to the change in the internal energy of the system.
\end{abstract}

\begin{keyword}
DFT \sep Electronic Structure \sep  Optical Properties  \sep Thermal Properties
\end{keyword}

\end{frontmatter}

\section{Introduction}

Low dimensional nanomaterials distinguish themselves by unique attributes, such as controllable properties and enhanced performance over bulk counterparts, and they contribute considerably to the development of nanotechnology. Among them, two-dimensional (2D) materials are extremely thin and have diverse physical characteristics. Scientists have been paying special attention to 2D materials as a unique group of materials after graphene's discovery \cite{novoselov2004electric}. 2D materials have shown considerable potential in future applications due to their unique physical properties related to their large surface area and extraordinary properties. This has prompted scientists to widen their investigation into 2D materials. Because of their unique electronic, optical, thermal, magnetic, mechanical, and catalytic properties, 2D materials are critical in science and technology, demanding extensive research \cite{novoselov2005two, gupta2015recent, shayeganfar2017monolayer, akinwande2017review, xiao2019atomic, abdullah2021properties, abdullah2023role}.

Following the discovery of graphene, a number of additional 2D materials are being studied for their novel features. Other 2D materials with characteristics equivalent to, or superior to, graphene can be synthesized and characterized \cite{mak2016photonics, bianco2013stability, sangwan2013low, vogt2012silicene, mannix2018borophene, zhu2015epitaxial}. Despite that MgC and MgO monolayers have yet to be experimentally widely tested, several groups have undertaken research to determine the presence of MgO monolayers in various material systems. A synthetic MgO nanosheet is created experimentally in several ways utilizing processes such as the layer-by-layer epitaxial growth mechanism generated using pulsed laser deposition \cite{matsuzaki2010layer}. MgO monolayers were created using a synthetic sol-gel method \cite{li2011experimental, kamarulzaman2016band, thomele2021cubes}, and the heat and pressure synthesis of methanol and ethanol from brown coal fly ash waste yields MgO monolayers \cite{zhu2012preparation, qian2021synthesis, liu2021ultrathin}. MgO monolayers are created by a wet chemical method involving chemical vapor deposition \cite{zhao2017generalized, wang2014mgo}.

A detailed analysis of multiple monolayers of the II-VI semiconductor family with features similar to 2D graphene have been performed using first principles calculation methodologies. Among them may be dynamically stable hexagonal monolayers such as MgO monolayers \cite{zheng2015monolayer, luo2019graphene}. However, monolayers with MgC and MgN structure characteristics are still in the theoretical modeling stage and have not been theoretically validated. There has been some research on them, but for different and novel structural forms such as MgN$_4$ nanosheets data indicates that an MgN$_4$ monolayer is dynamically and thermally stable. In the electronic structure of MgN$_4$ nanosheets, anisotropic Dirac cones have been identified \cite{mortazavi2021ultrahigh}, the carrier mobility of the MgN$_2$ monolayer was 104 cm$^2$V$^{-1}$s$^{-1}$, and an indirect band gap was found to be 2.33 eV. Further significantly, the MgN$_2$ monolayer covers the reduction levels necessary for water splitting, making it a workable material for hydrogen production \cite{wei2018promising}. However, a number of studies on the MgO monolayer point to their remarkable surface activity. MgO monolayers offer a variety of fascinating uses, such as MgO(111) for methanol-based fuel cells \cite{hu2007mgo}, having the potential to trigger unusual properties \cite{liu2021atomically}, catalytic action to prevent adsorption of certain toxic environmental pollutants \cite{zhu2006efficient}, catalytic activity of carbon dioxide and methane conversion \cite{lin2015effect}. The have been used for the development of highly energy-dense, reversible rechargeable batteries with exceptional capacity retention at high current densities for ion batteries as anode materials \cite{zhou2020redistributing}, as well as offering an appropriate substrate \cite{smerieri2015spontaneous, zhang2014carbon}. Additionally, a novel 2D Mg$_2$C nanosheet with quasi-planar hexa-coordinate magnesium and carbon has recently been the subject of theoretical research for electronic structure calculations. This monolayer can under biaxial tensile strain switch between being a metal and a semiconductor as simulations using density functional theory and molecular dynamics indicate \cite{meng2017metal}.

On the other hand, despite the fact that there have been several theoretical analyses of MgO nanosheets for a significant number of specified parameters, further studies are still necessary. For example, The structure, the stability, and the electronic properties of MgO nanosheets have been investigated using DFT. These nanosheets have been found to have semiconductive characteristics, with band gaps that differ based on strain from 4.23 to 4.38 eV \cite{zhang2012structural}. The band gap for MgO monolayers with a direct gap is 3.1 eV with GGA and 4.2 eV with GGA-mBJ. Various optical properties that demonstrate the semiconducting character of MgO nanosheets are determined using the RPA approximation \cite{nourozi2019electronic}. The phonon and the thermodynamic properties of MgO (111) and MgO (100) nanosheets have been evaluated using DFT, and it was found that MgO (100) displays higher dynamical stability than MgO (111) \cite{yeganeh2019vibrational}. A Boltzmann transport approach and first principles study have been used to examine how quantum confinement affects the thermoelectric and the electronic characteristics of both MgO and (111) \cite{asl2022two}. The electronic and the optical properties of a MgO nanosheet under compressive and tensile strain have been investigated using DFT. The band gap energy is 3.45 eV and reduces with tensile strain, while a significant increase is shown when compressive strain is applied \cite{yeganeh2019effects}.

Theoretical studies of a doped MgO monolayer by a particular element suggest potential applications for the future with varying characteristics. For example, findings from investigation of the electronic and the magnetic properties of a transition metal doped MgO monolayer suggest that the band gaps and the magnetic characteristics of MgO nanosheets can be changed \cite{wu2014electronic}. The band gaps can be controlled by B, C, N, and F doping of MgO nanosheets and they are semiconductors according to DFT due to impurity states emerging inside the band gap, whereas an F-doped MgO nanosheet makes the transition from a semiconductor to a metal \cite{wu2016first, moghadam2018electronic}. The effects of point defects and doping by Si and Ge on the electronic and the magnetic properties of a MgO monolayer has been studied. Band gap fluctuations can cause the material to change from a non-magnetic material to a ferromagnetic semiconductor under certain cases \cite{van2022defective}.

We concentrate in this work on the electronic, the thermal, and the optical characteristics of hexagonal MgC, MgN, and MgO monolayers within the concept of DFT modeling due to its enormous potential to advance our knowledge. The electronic, the thermal, and the optical characteristics of MgC and MgN are examined for the first time in this paper. Additionally, we look at additional research on various physical properties of MgO monolayers. The results for these monolayers are compared to one another and to the data that is available for the MgO monolayer. The findings demonstrate different electronic, thermal, and optical characteristics, and suggest that MgC, MgN, and MgO monolayers are important in thermal and optoelectronic applications \cite{abdullah2020electronic}.
It is anticipated that functionalizing the MgX monolayer leads to enhancement of its optoelectronic properties, making it more suitable for practical applications in nano and spintronic devices \cite{VANON2022106876}. In addition, the MgX can also be used to improve the phase sensitivity of sensors \cite{8933899, 8049273}.

The paper is structured as follows:  \sec{Methodology} contains information about the computational methods, and \sec{Results} shows the computed electronic, thermal, and optical characteristics for a MgX monolayer. The section \sec{conclusion} of the work contains the paper's conclusion.

\section{Methodology}\label{Methodology}

We assume a 2D hexagonal h-MgX monolayer made up of $2\times 2$ supercells with an equal amount of Mg and X (where X=C, N, and O) atoms. The cut-off for the kinetic energy and the charge densities are considered to be $1088.5$~eV, and $1.088 \times 10^{4}$~eV, respectively. The MgX monolayers are fully relaxed with the stated values when all the forces are below $10^{-5}$ eV/$\angstrom$ using a Monkhorst-Pack grid of $18 \times 18 \times 1$ sites. The vacuum layer or the inter-layer distance is $20 \, \angstrom$ in the $z$-direction eliminating the interaction of the MgX monolayers \cite{ABDULLAH2022106409}.

In the context of DFT, the generalized gradient approximation (GGA) and the Perdew-Burke-Ernzerhof (PBE) functional are used to approximate the exchange and the correlation terms \cite{Giannozzi_2009, giannozzi2017advanced}. The Kohn-Sham density functional theory provides the basis for the QE
(Quantum Expresso) package. The band structure and the density of states (DOS) are calculated using the Self-Consistent Field (SCF) and the Non-Self-Consistent Field (NSCF) methods, respectively. We utilize a Monkhorst-Pack grid of $18 \times 18 \times 1$ for the SCF  and $100 \times 100 \times 1$ for the NSCF for these computations \cite{ABDULLAH2022115554}. QE is also used to determine the optical properties of the MgX monolayers with an optical broadening of $0.1$~eV \cite{ABDULLAH2022114590}.

Ab-initio molecular dynamics, AIMD, simulations are done to investigate the thermal stability of
a MgX monolayer by using a $2\times2\times1$ super-cell at room temperature, $300$ K, with total simulation time of $5$ ps with $1$~fs time steps.

\section{Results} \label{Results}

The findings on the electronic, the thermal, and the optical properties of the h-MgX monolayers are presented in this section.

\subsection{Electronic Properties}
In this section, the electronic characteristics of h-MgX monolayers are explored using DFT. The MgC, the MgN, and the MgO monolayers have a hexagonal structure with almost no planar buckling in their flat forms because of the strong $\sigma\text{-}\sigma$ bonds produced by a sp$^2$ hybridization.
In order to evaluate the stability of the structures we turn first to the energy required to build a monolayer structure, the formation energy. The formation energies of the MgC, MgN, and MgO monolayers are found to be $1.51$, $1.39$ and $1.24$~eV, respectively. A lower formation energy can indicate a stable monolayer structure. Using the results of our DFT simulations and the formation energies of the three different MgX monolayer forms, we can assign to them the following statuses: MgC \textgreater\  MgN \textgreater\  MgO. In comparison to the monolayers of MgC and MgN being studied, the MgO monolayer is the most energetically stable structure \cite{PhysRevB.92.115307}.

\begin{figure}[htb]
	\centering
	\includegraphics[width=0.45\textwidth]{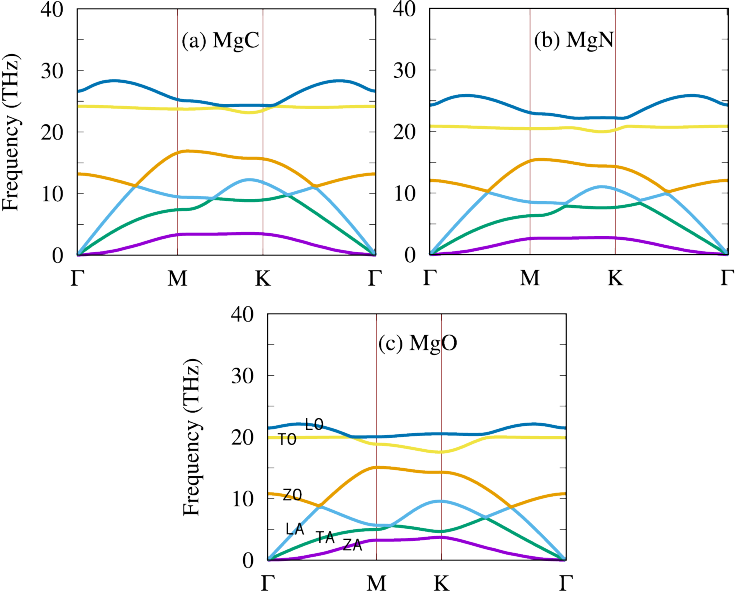}
	\caption{Phonon band structure of the (a) MgC, (b) MgN, and (c) MgO monolayers.}
	\label{fig01}
\end{figure}

The dynamical stability of the MgX monolayers can be checked by calculating the phonon band structure as
is presented in \fig{fig01}.
It is known that the $k \rightarrow 0$ phonon dispersion of the LA and the TA branches is linear, but that of the ZA branch due to out of plane acoustical modes is quadratic since transversal forces decay rapidly.  Among these, the LA and TA phonon branches are the heat-carrying modes.
It has been demonstrated that the bending branch ZA makes negligible contribution to
the thermal conductivity \cite{klemens2001theory}. It has been reported that if the ZA branch becomes soft
and imaginary frequencies appear in calculations in the BZ for a monolayer, the structure becomes unstable.
This phenomena is interpreted as the instability against long-wavelength transversal waves, while
a positive frequency of ZA shows dynamic stability of the monolayer. We can confirm that all three considered monolayers here are dynamically stable as no imaginary frequencies emerge in the calculations.

The geometrical characteristics of pure MgX monolayers are what we first look at. The entire thickness of the MgX monolayer, one atom, is provided by the Mg-C, Mg-N, and Mg-O bonds. The buckling characteristics in the flat form of these monolayers are slightly different from zero due to variation of the atomic diameters, which is an essential point to identify. Similar to graphene, these monolayers feature flat hexagonal honeycomb structures. The most energetically advantageous arrangements of the three monolayers may very well be derived from the bond lengths. The stability of a monolayer is significantly linked with the bond lengths. The findings of this study are novel for the MgC and the MgN monolayers, and the bond lengths for pure Mg-C and Mg-N monolayers are $2.13$~$\angstrom$ and $2.012$~$\angstrom$ and the lattice constants are $3.7$ and $3.48$~$\angstrom$, respectively. The Mg-O bond length and the lattice constant for a pure MgO monolayer are further shown via PBE optimization to be $1.89$ $\angstrom$ and $3.3$ $\angstrom$, respectively. These data are consistent with recent theoretical investigations of MgO monolayers \cite{wu2016first, nourozi2019electronic}.

\begin{figure}[htb]
	\centering
	\includegraphics[width=0.45\textwidth]{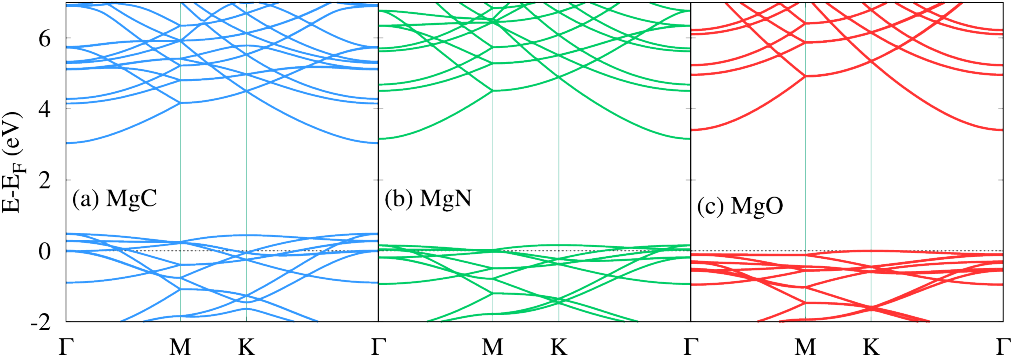}
	\caption{Electronic band structure of the (a) MgC, (b) MgN, and (c) MgO monolayers.}
	\label{fig02}
\end{figure}

The electronic bandstructure of the planar configuration of the MgX monolayers' is seen in \fig{fig02} and they all exhibit specific features. According to the results, which are in agreement with previous results \cite{wu2014electronic, nourozi2019electronic}, MgO is semiconducting monolayer with an indirect band gap of $3.4$ eV. However, the band structures of the hexagonal MgC and MgN monolayers demonstrate that the Fermi energy minimally crosses the valence band area, causing metallic properties of these two monolayers. This suggests that the type of crystal structure is essential for determining how a material will eventually take on its phase.

\begin{figure}[htb]
	\centering
	\includegraphics[width=0.45\textwidth]{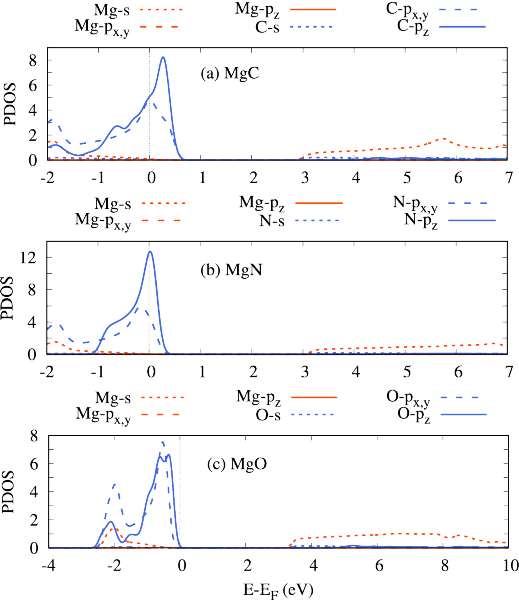}
	\caption{Partial density of states of the (a) MgC, (b) MgN, and (c) MgO monolayers.}
	\label{fig03}
\end{figure}

It is possible to study the atomic orbital contribution to the band structures by looking at the partial density of states of each of these three monolayers, this is done in \fig{fig03}. The atoms' $p$-orbital contribution exhibits isotropic distribution along the $x$ and $y$-axis, to start.
It means that the physical characteristics must be sensitive to symmetries in the $xy$-plane. The valence bands are created via a weak hybridization between the Mg ($p$-orbital) and the O ($p$-orbital), with the O atoms supplying the larger contribution. The Mg ($s$-orbital) is responsible for the majority contribution to the conduction band, according to the PDOS of a MgO monolayer. The PDOS of both the MgC and the MgN monolayers shows little variation to the situation in the MgO monolayer, dispite that the N atoms' $p$-orbitals show a slightly reduced contribution. Additionally, there is a little shift in the Fermi energy that produces the metallic characteristic, due to the changed distribution of the electron densities, and this change is not as significant for MgN as it is for MgC.

\subsection{Thermal Properties}
\begin{figure}[htb]
	\centering
	\includegraphics[width=0.5\textwidth]{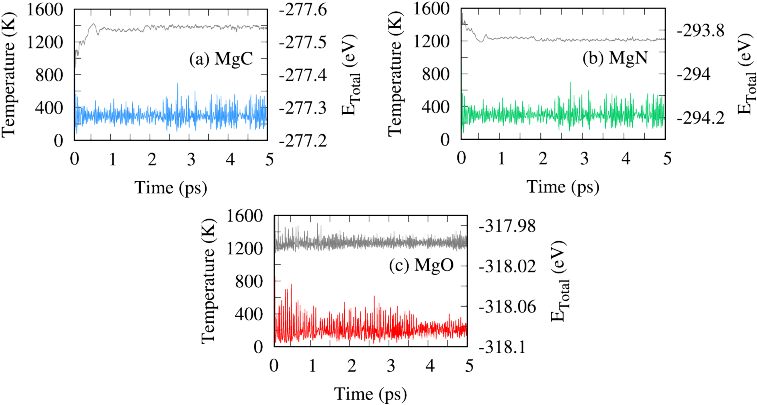}
	\caption{Temperature versus the AIMD simulation time steps at $300$ K for optimized MgC (a), MgN (b), and MgO (c). The gray solid line is the variation of total energy with time for all considered monolayers.}
	\label{fig04}
\end{figure}

The thermal stability of the MgX monolayers is tested for approximately $5$ ps with a time step of $1.0$~fs as is presented in \fig{fig04}. The temperature curve of the MgX monolayers neither displays large fluctuations in the temperature nor serious structure disruptions or bond breaking at $300$~K. This indicates that the MgX monolayers are thermodynamically stable structures. In addition, the variation of total energy (gray solid line) per atom is less than $1.0$ eV, which is in the acceptable range similar to many studies in the literature \cite{D1CP01183A} indicating the absence of large energy fluctuation.

\begin{figure}[htb]
	\centering
	\includegraphics[width=0.35\textwidth]{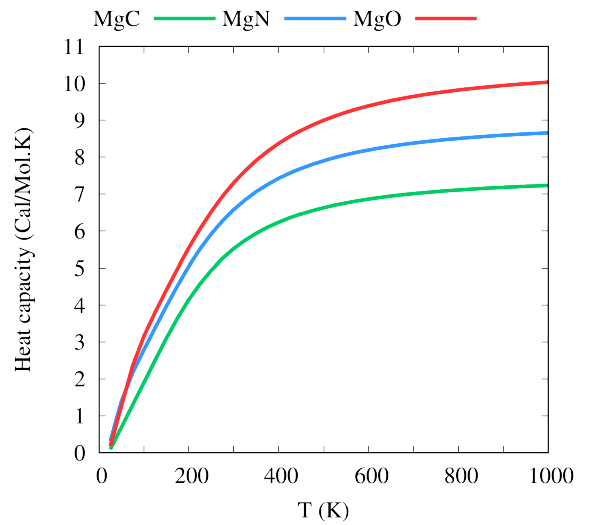}
	\caption{Heat capacity versus temperature for MgC (green), MgN (blue), and MgO (red) monolayers.}
	\label{fig05}
\end{figure}

The heat capacity for the MgX monolayers' is shown in \fig{fig05}. The MgX monolayer has outstanding stability at high temperatures, as was shown by the temperature versus time curve in \fig{fig04}.
The quantity of heat that must be applied to a material in order to cause a unit change in temperature is known as a material's heat capacity. The change in the heat capacity starts at low temperatures, and as the temperature rises, the heat capacity starts to increase. All three MgX monolayers exhibit higher heat capacity with increasing temperatures as more phonons are activated and the internal energy rises quickly. Since so many energy levels are occupied at high temperatures, the rate of increase stabilizes, as is shown in \fig{fig05}. The increase ratio in the heat capacity is greater for the MgO monolayer compared to the MgC and the MgO monolayers.
This can be explained by the classical theory that expects higher heat capacity for the systems with stronger bonds \cite{MORTAZAVI2021100257}.
\begin{figure*}[htb]
	\centering
	\includegraphics[width=0.9\textwidth]{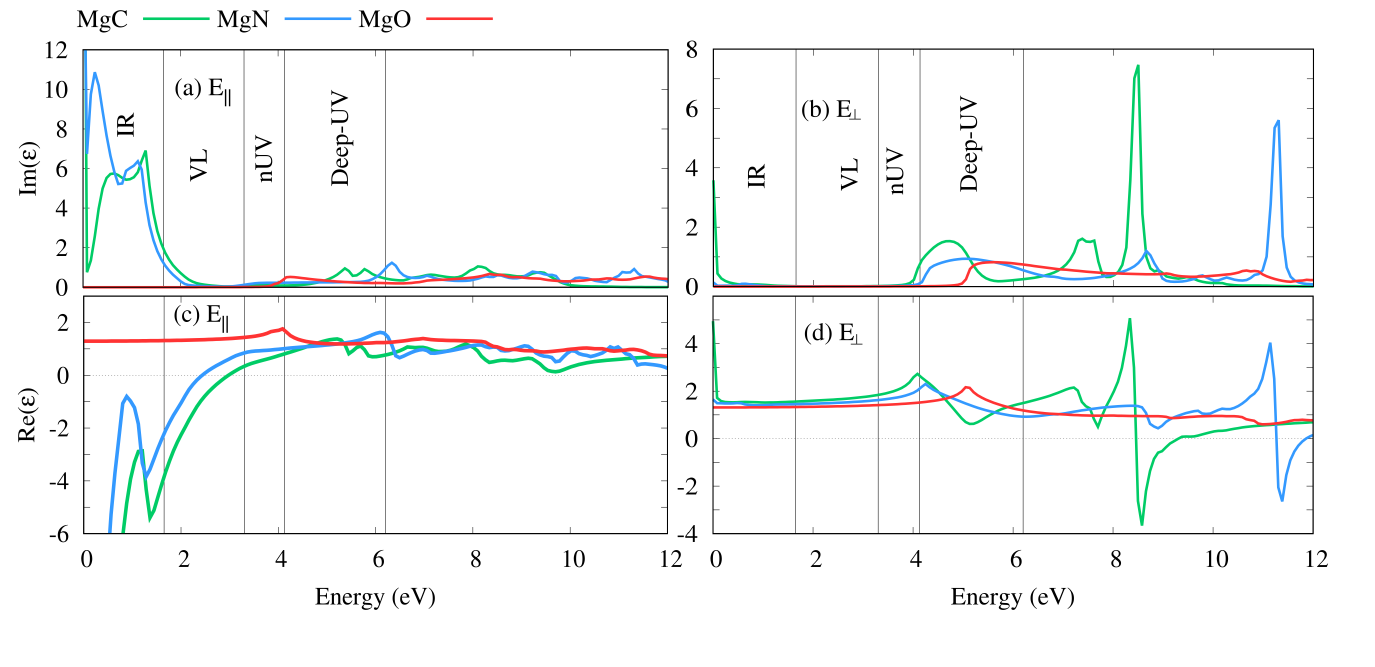}
	\caption{The Im$(\varepsilon)$ (a,b) and the Re$(\varepsilon)$ (c,d) presented for MgX           monolayers in the incident light polarized along the $x$ (E$_{\parallel}$), and $z$ (E$_{\perp}$) directions.}
	\label{fig06}
\end{figure*}
A strong bond is obtained if the electronegativity difference across a bond is high. The average electronegativity difference across the bonds is 2.13 (MgO), 1.73 (MgN), and 1.24 (MgC).
So, the heat capacity for MgO has the highest value among all three considered monolayers as it has the strongest bond with high electronagetivity across it.

\subsection{Optical Properties}
The complex dielectric function illustrates how materials react to electromagnetic radiation. In addition, the dielectric constant of a substance can be defined as a measure of its ability to store electrical energy. It is an expression of the extent to which a material holds or concentrates electric flux. We use the random phase approximation (RPA), which employs a very dense mesh grid of $100 \times 100 \times 1$ in the Brillion zone to get precise results \cite{ABDULLAH2022106835, abdullah2022electronic}. We first take into account the real and the imaginary components of the dielectric function for the parallel E$_{\parallel}$ and perpendicular E$_{\perp}$ directions of the incoming electric field shown in \fig{fig06} \cite{ABDULLAH2023116147}. The locations of the infrared, IR, ($0\text{-}1.8$~eV), the visible regime, VL, ($1.8\text{-}3.3$~eV), the near ultraviolet, nUV, ($3.3\text{-}4.13$~eV), and Deep-UV ($4.13\text{-}6.2$~eV) of the electromagnetic spectrum are identified.

The optical properties of the crystalline monolayers are significantly influenced by the Im($\varepsilon$). It is noticeable that these optical characteristics for the MgX monolayers are anisotropic with respect to the polarization directions. The optical band gap transition for the MgO monolayers in the $E_{\parallel}$ case is illustrated by a tiny peak in Im($\varepsilon$) arising at $4.2$~eV in the nUV region, confirming the monolayer's semiconductor property. However, the peak is seen in the deep-UV region for E$_{\perp}$ at $5.1$~eV. This confirms that the MgO monolayer is active in the nUV and the Deep-UV region for different directions of electric field. In comparison, the IR region of MgC (green) and MgN (green) exhibits sharp peaks in the Im($\varepsilon$). The intraband transition is caused by the crossing of the Fermi energy through the valence bands. These IR peaks are ineffective because metallic materials are not interesting for optoelectronic devices. Interesting physical aspects such as the static dielectric constant and the plasmon energy are present in the real component of the dielectric function. The value of the real component of the dielectric function when the energy of incoming electric or photon field approaches zero is known as the static dielectric constant. In the $E_{\parallel}$ and the $E_{\perp}$ directions, MgO monolayer has a static dielectric constant of $1.16$ and $1.19$, respectively. This supports the MgO's role as a semiconductor. The metallic nature of the two monolayers, MgC and MgN for $E_{\parallel}$ have different static dielectric values, due of the anisotropic nature of Re($\varepsilon$), and $E_{\perp}$ has a static dielectric constant $1.66$ for MgN, whereas for MgC, it is $4.94$, a rather high value.

\begin{figure*}[htb]
	\centering
	\includegraphics[width=0.8\textwidth]{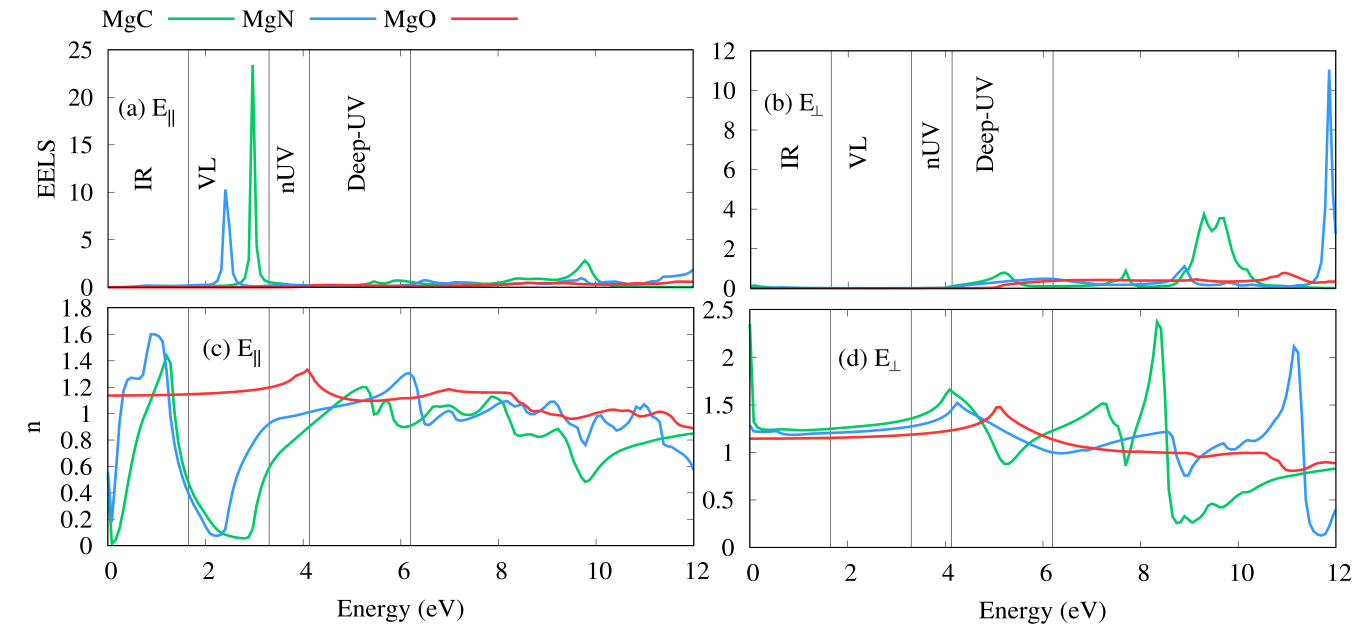}
	\caption{Electron energy loss spectrum, EELS, and refractive index, $n$, for MgX  monolayers in the cases of (E$_{\parallel}$), and (E$_{\perp}$).}
	\label{fig07}
\end{figure*}

\begin{figure*}[htb]
	\centering
	\includegraphics[width=0.8\textwidth]{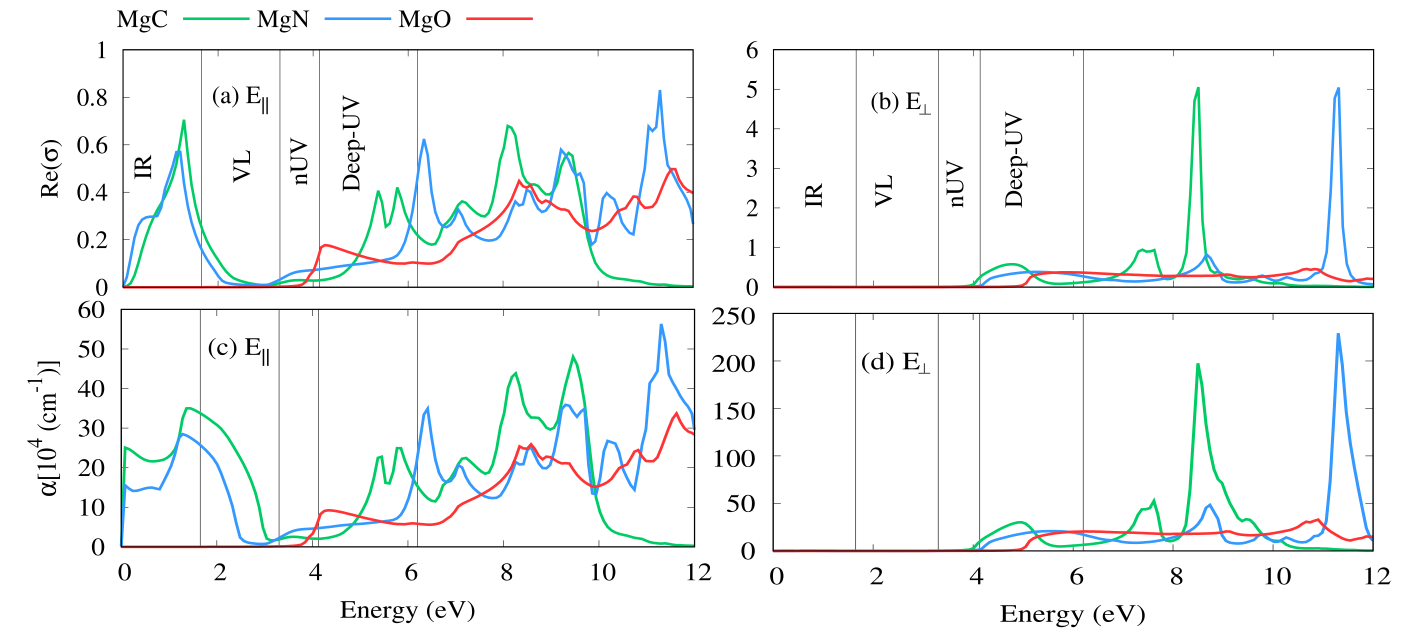}
	\caption{Re($\sigma_{\rm optical}$) and  $\alpha$ in the cases of  $E_{\parallel}$, and  $E_{\perp}$ to the MgX monolayers.}
	\label{fig08}
\end{figure*}
Studying the refractive index, $n$, and the electron energy loss spectra, EELS, is informative for both E$_{\parallel}$ and E$_{\perp}$, as is shown in \fig{fig07}. The energy loss function diagram provides important details regarding the optical parameters during the quickly moving electron's interaction with the material. Both the MgC and the MgN monolayers exhibit a metallic property for the E$_{\parallel}$ polarization to the monolayers, as is seen by the first energy loss function peak in the visible area. This implies that interband transition in the monolayers have a strong ability to absorb visible light. There are peaks for MgC and MgN in the Deep-UV region in the case of the E$_{\perp}$ polarization due to intraband transitions indicating that the monolayers in the high energy range have a great capacity to absorb light. On the other hand, the MgO monolayer energy loss function diagram shows no noticeable spectrum peak for both polarizations. In the case of E$_{\parallel}$, the maximum refraction index 1.3 of a monolayer of MgO is seen at 4.1 eV, while there are variations in the refraction in the infrared and the visible spectrum for MgC and MgN monolayers. On the other hand, the peak in E$_{\perp}$ moves to the Deep-UV region for the MgO monolayer. Furthermore, the fluctuation of the peaks of the refractive index indicate plasmon’s fluctuations for both the MgC and the MgN monolayers at low energies. Incident photons cause collective vibrations of the electrons on the surface, which is compatible with the minimum of the real part of the dielectric function Re$(\varepsilon)$. A metallic behavior is demonstrated in this region.

Finally, the real part of the optical conductivity, Re($\sigma_{\rm optical}$), and the absorption coefficient, $\alpha$, of MgX are displayed in \fig{fig08} for both $E_{\parallel}$ and $E_{\perp}$. Since the MgO monolayer has a semiconducting character, the first peak in these spectra is seen for both the Re($\sigma_{\rm optical}$) and $\alpha$ spectra in the nUV region in the case of $E_{\parallel}$. However, due to the semimetal structure of MgC and MgN, there is no interesting peak available. In the case of $E_{\perp}$, the MgO monolayer has the first peaks for both Re($\sigma_{\rm optical}$) and $\alpha$ in the deep-UV area of energy, whereas the peaks for the MgC and MgN monolayers exist in the n-UV region. In contrast, MgC and MgN exhibit two intensive peaks in the deep-UV range because of intraband transitions and a different band structure compared to an MgO monolayer.

\section{Conclusions}\label{conclusion}
The electronic, the structural, the thermal, and the optical properties of a hexagonal pristine MgX monolayer are investigated using first-principles calculations (where X is C, N, and O).
The AIMD simulations and the phonon band structure calculations confirm that the MgX monolayers all have a thermodynamic and a dynamic stability. The band structure and density of states calculations show that the MgC and the MgN have metallic characters, while the MgO monolayer has a semiconducting property.
The electronegativity differential across the Mg-X bond determines whether the MgX monolayers have active thermal properties or not. It is seen that the MgO monolayer has a high heat capacity compared to the MgC and the MgN monolayers as the Mg-O bond has a high electronegativity resulting from the strong Mg-O bonds.
A study of the optical properties of the MgX monolayers indicates that the MgO semiconducting monolayer exhibits an active optical response in the n-UV.

\section{Acknowledgment}
The University of Sulaimani and the Research Center of Komar University of Science and Technology provided financial assistance for this work.
The Division of Computational Nanoscience at the University of Sulaimani supplied the resources used for the computations.



\end{document}